# General Two-Parameter Model of Alpha-Relaxation in Glasses

Valeriy V. Ginzburg[1,*], Oleg Gendelman[2], Riccardo Casalini[3], and Alessio Zaccone[4]

[1]Department of Chemical Engineering and Materials Science, Michigan State University, East Lansing, Michigan, USA 48824

[2]Faculty of Mechanical Engineering, Technion, Haifa 32000003, Israel

[3]Chemistry Division, Naval Research Laboratory, 4555 Overlook Avenue SW, Washington, D.C., USA 20375

[4]University of Milan, Department of Physics, via Celoria 16, 20133 Milano, Italy

*Corresponding author: ginzbur7@msu.edu

**ABSTRACT**. In the vicinity of the glass transition, the characteristic relaxation time (e.g., the $\alpha$-relaxation time in dielectric spectroscopy) of a glass-former exhibits a strongly super-Arrhenius temperature dependence, as compared to the classical Arrhenius behavior at high temperatures. A comprehensive description of both regions thus requires five parameters. Here, we demonstrate that many glass-formers exhibit a general scaling, with only two material-specific parameters setting the timescale and the temperature scale; the other three being material-independent constants. Furthermore, we show that the master curve can be described by the recently developed two-state, two-(time) scale (TS2) theory (Soft Matter **2020**, *16*, 810) and regress the general TS2 parameters. We also show the connection between the TS2 model and the Hall-Wolynes elastic relaxation theory.

*Introduction.* Glass transition is a phenomenon ubiquitous for many classes of amorphous materials (networks, colloids, molecular fluids, polymers, etc.) [1–12] One of its features is the super-Arrhenius dependence of the $\alpha$-relaxation time or viscosity on the temperature. This behavior is often described using the Vogel-Fulcher-Tammann-Hesse [13–15] (VFTH) equation, $\log(\tau[T]) = \log(\tau_\infty) + D/(T - T_0)$, and usually explained on the basis of vanishing configurational entropy [16] or vanishing free volume. [17] Notably, the VFTH equation predicts that the relaxation time diverges as $T \rightarrow T_0$. This divergence has been questioned by many authors (see, e.g., [18–21]) and several non-divergent models have also been proposed. [22–25] The super-Arrhenius behavior is described by multiple theoretical frameworks, such as dynamic facilitation [26–30], elastic models [8–10], random first-order theory (RFOT) [31–33], mode-coupling theory (MCT) [34–37], elastically collective nonlinear Langevin equation (ECNLE) [38–41], and many others. By combining the relaxation time dynamics with equation of state, one can simultaneously model the dielectric relaxation and the volumetric or calorimetric experiments, see, e.g., Douglas *et al.* [42–44] and Lipson *et al.* [45–47].

Recently, we proposed a simple theoretical framework, the "two-state, two-(time)scale" (TS2) model that is able to capture the relaxation time behavior and describe some other experiments (volume relaxation, stress relaxation, density changes on cooling, etc.). [24,48–52] Yet one challenge was to find a simple and robust model parameterization. Here, we use historical data for a number of network, molecular, and polymeric glasses to demonstrate the broad validity of the TS2 relaxation time dependence and to regress the material-independent TS2 parameters.

*The relaxation master curve.* In Figure 1, Angell plots [1] are shown for the viscosities of selected molecular and network glasses (*a*) and dielectric $\alpha$-relaxation times of selected polymer glasses (*b*). The slope of the curve at the point $T_g/T = 1$ is termed fragility, *m,*

$$m = \left.\frac{d \log Y}{d\left(\frac{T_g}{T}\right)}\right|_{\frac{T_g}{T}=1} \tag{1}$$

Here, $Y = \eta$ or $\tau_\alpha$, and $T_g$ is defined as the temperature for which $\eta = 10^{12}$ Pa·s or $\tau_\alpha = \tau_g = 10^2$ s. Most polymers have fragilities in the range ~50 - 200 [53] (Figure 1b). Among the molecular and network glasses, silica has the lowest fragility (m ~ 20) and exhibits Arrhenius behavior; other non-polymeric glass-formers have fragilities in a very broad range. The temperature dependence of the relaxation time in the VFTH regime can be conveniently written as,

$$\log\left(\frac{\tau_\alpha[T]}{\tau_\infty}\right) = \frac{K^2}{K + m\left(\frac{T}{T_g} - 1\right)} \tag{2}$$

where $K = \log(\tau_g/\tau_\infty)$, and $\tau_\infty = 10^{-13}$–$10^{-11}$ s. [54–56]

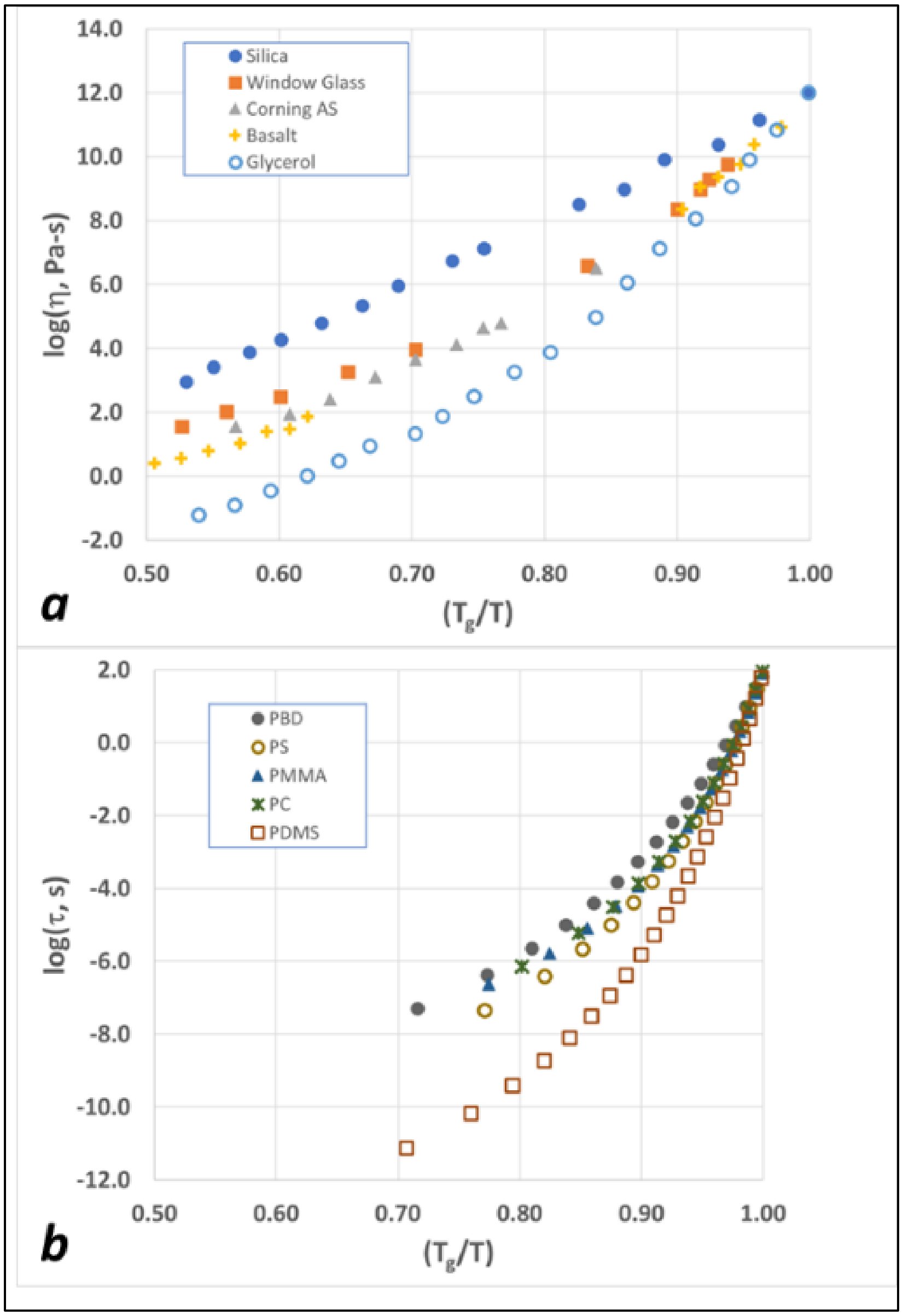

*Figure 1. Angell plots: (a) viscosities of several network and molecular glasses (data from Ref. [23]); (b) $\alpha$-relaxation times of several polymers (data from Refs. [54,57,58]). Abbreviations: Corning AS – Corning alumosilicate glass; PBD – polybutadiene; PS – polystyrene; PMMA – poly(methylmethacrylate); PC – bisphenol-A polycarbonate; PDMS – poly(dimethyl siloxane).*

The Blochowicz equation (2) has three parameters ($\tau_\infty$, $m$, and $T_g$). It generally does not describe strong glasses or the high-temperature Arrhenius region for fragile glasses and predicts the divergence of the relaxation time as $T \to T_0 = T_g(1 - K/m)$.

To develop a phenomenological general description of both strong and fragile glasses in both Arrhenius and VFTH regions, we utilize the combined vertical and horizontal shift procedure similar to the one outlined by Bailly et al. [59] The temperature axis is first converted to logarithmic scale, and then horizontal and vertical shifts are applied to collapse all curves onto a single master curve. (Here, the horizonal shift corresponds to $\ln(T_x/T_g)$ and the vertical shift is $\log(\tau_g/\tau_{el})$; the meaning of the new temperature parameter $T_x$ and time parameter $\tau_{el}$ will be discussed below). For our analysis, we use the data for 35 glass-formers, taken from Refs. [54,57,58,60–64] (polymers) and [23,65] (network and molecular glasses); see Supplementary Information Table S1 [66] for details. The results for one subset containing 15 materials (3 network glasses, 9 polymer glasses, and 3 molecular glasses) are shown in Figure 2a (log-log plot) and Figure 2b (Arrhenius plot); the other subsets are given in Supplementary Information [66] as Figures S1 – S3. The data are analyzed in the temperature range between $T_g$ and about $1.4T_g$; the typical temperature for the VFTH-to-Arrhenius transition for polymers is about $1.2T_g$, [67] so our model probes both VFTH and Arrhenius regions. However, it is important to note that some experiments, e.g., on propylene carbonate (see Schönhals et al. [68]) and ortho-terphenyl (see Caruthers and Medvedev [65] and references therein) explore the Arrhenius region at much higher temperatures and observe the activation energies significantly smaller than the Wang-Porter [69] prediction. This might be due to some additional transitions within the liquid state.

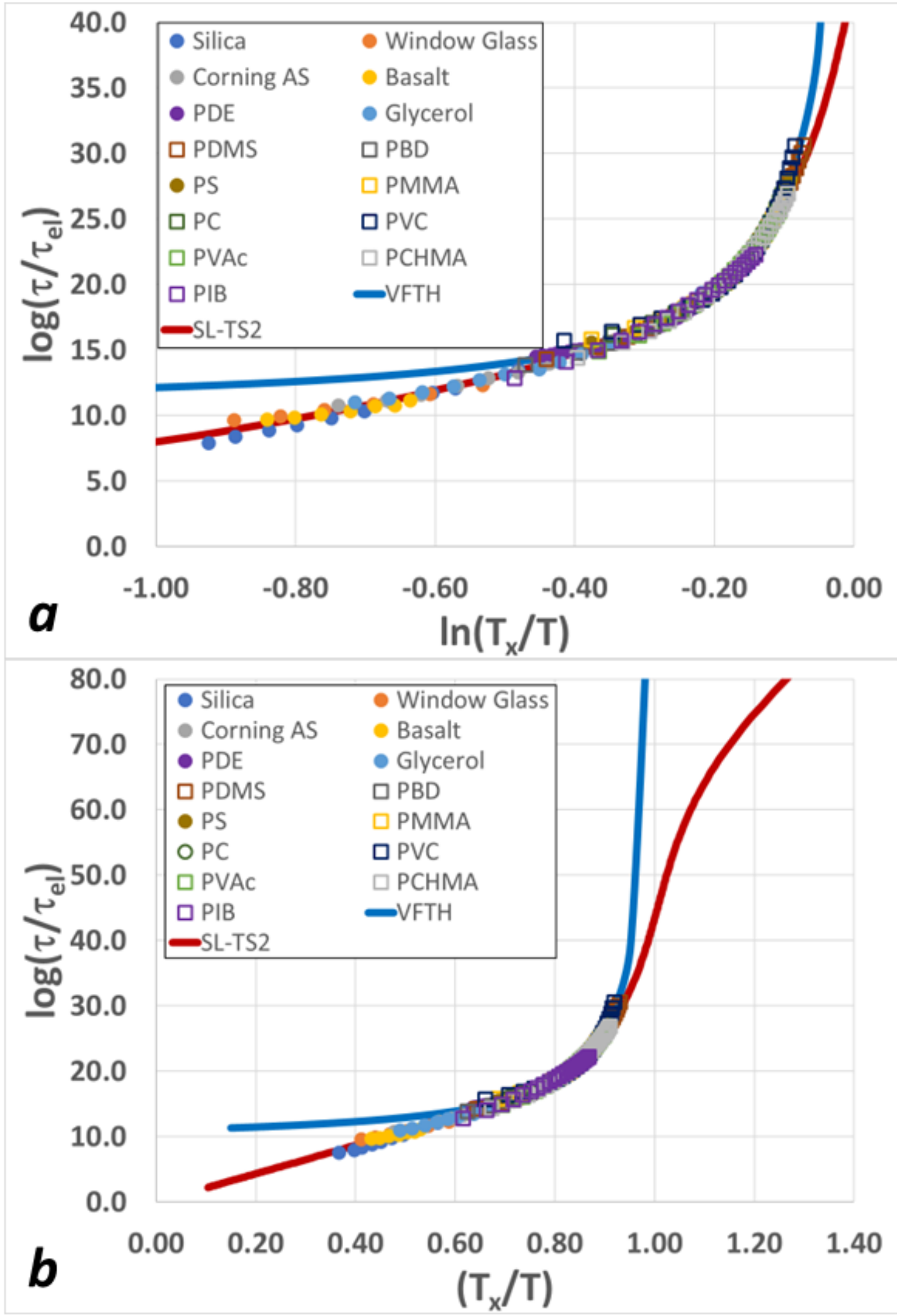

*Figure 2. The scaling for the $\alpha$-relaxation and viscosity data for 15 glass-formers. (a) Log-log plot, with details about all the glass-formers used in the modeling. (b) Linear-log plot for the same systems, but with expanded temperature range, extrapolating into the glassy region. The red line corresponds to theoretical SL-TS2 model described below; the blue line is the VFTH fit. Abbreviations: Corning AS – Corning aluminosilicate glass; PDMS – poly(dimethylsiloxane); PBD – polybutadiene; PS – polystyrene; PMMA – poly(methyl methacrylate); PC – bisphenol-A-polycarbonate; PVC – poly(vinylchloride); PDE -- phenylphthalein-dimethylether; PVAc – poly(vinylacetate); PCHMA – poly(cyclohexyl methacrylate); PIB – poly(isobutylene).*

On the log-log plot (Figure 2a), after the vertical and horizontal shifts, notwithstanding the large difference of the glass formers' fragility, the data collapse onto a single master curve with an Arrhenius region on the left (high temperatures) and a VFTH region on the right (temperatures near $T_g$). The reference is selected so that when the data are re-plotted in the Arrhenius coordinates (Figure 2b), the intercept is set to equal to zero. The resulting master curve is successfully (standard deviation of log( $\tau$) less than 1.0, see Table S1 [66]) fit to the TS2 function,

$$\log\left(\frac{\tau_\alpha[X]}{\tau_{el}}\right) = \frac{\overline{E}_1}{X} + \frac{\overline{E}_2 - \overline{E}_1}{X}\psi[X] \tag{3}$$

$$\psi[X] = \left\{1 + \exp\left[\frac{\Delta S}{k_B}(1 - X^{-1})\right]\right\}^{-1} \tag{4}$$

Here, $k_B$ is the Boltzmann's constant, $X = T/T_X$ is the reduced temperature, $\overline{E}_1 = E_1/k_B T_X$ , and $\overline{E}_2 = E_2/k_B T_X$ are the reduced activation energies of the "liquid" and "solid" states, respectively, and $\Delta S/k_B$ is the entropy difference between the two states. The timescale $\tau_{el}$ is an intrinsic characteristic of the material, although its physical meaning is somewhat uncertain. Equation 4 can be also thought of as an approximate analytical description of the numerical solution to the more fundamental Sanchez-Lacombe two-state, two-(time)scale (SL-TS2) model,

$$\beta G = \left[\frac{r}{\nu}\right]\left[\left(\nu\frac{\psi}{r}\right)\ln\left(\nu\frac{\psi r_S}{r}\right) + \left(\nu\frac{1-\psi}{r}\right)\ln\left(\nu\frac{(1-\psi)r_L}{r}\right) + (1-\nu)\ln(1-\nu)\right]$$
$$-\frac{\overline{\varepsilon^*}}{X}\left[\frac{r}{\nu}\right]\left[\left(\nu\frac{\psi r_S}{r}\right)^2 + 2\alpha_{SL}\left(\nu\frac{\psi r_S}{r}\right)\left(\nu\frac{(1-\psi)r_L}{r}\right) + \alpha_{LL}\left(\nu\frac{(1-\psi)r_L}{r}\right)^2\right]$$

(5)

Here, $G$ is the Gibbs free energy, $\nu$ is the "occupancy", z is the lattice coordination number, $r_S$ and $r_L$ are the numbers of lattice sites occupied by the "solid" and "liquid" clusters, respectively; $r = \psi r_S + (1-\psi)r_L$ is the average size of the cooperatively rearranging region (CRR). The normalized energy parameter, $\overline{\varepsilon^*}$, describes the S-S interaction, and the dimensionless parameters $\alpha_{LL}$ and $\alpha_{SL} = \sqrt{\alpha_{LL}}$ describe the relative strength of the L-L and S-L interactions relative to the S-S one. The material-independent TS2 and SL-TS2 parameters are summarized in Supplementary Information Table S-II [66]. The liquid-state activation energy parameter, $\overline{E}_1$, agrees well with the estimates of Wang and Porter [69] - see Supplementary Information [66] for more detailed discussion.

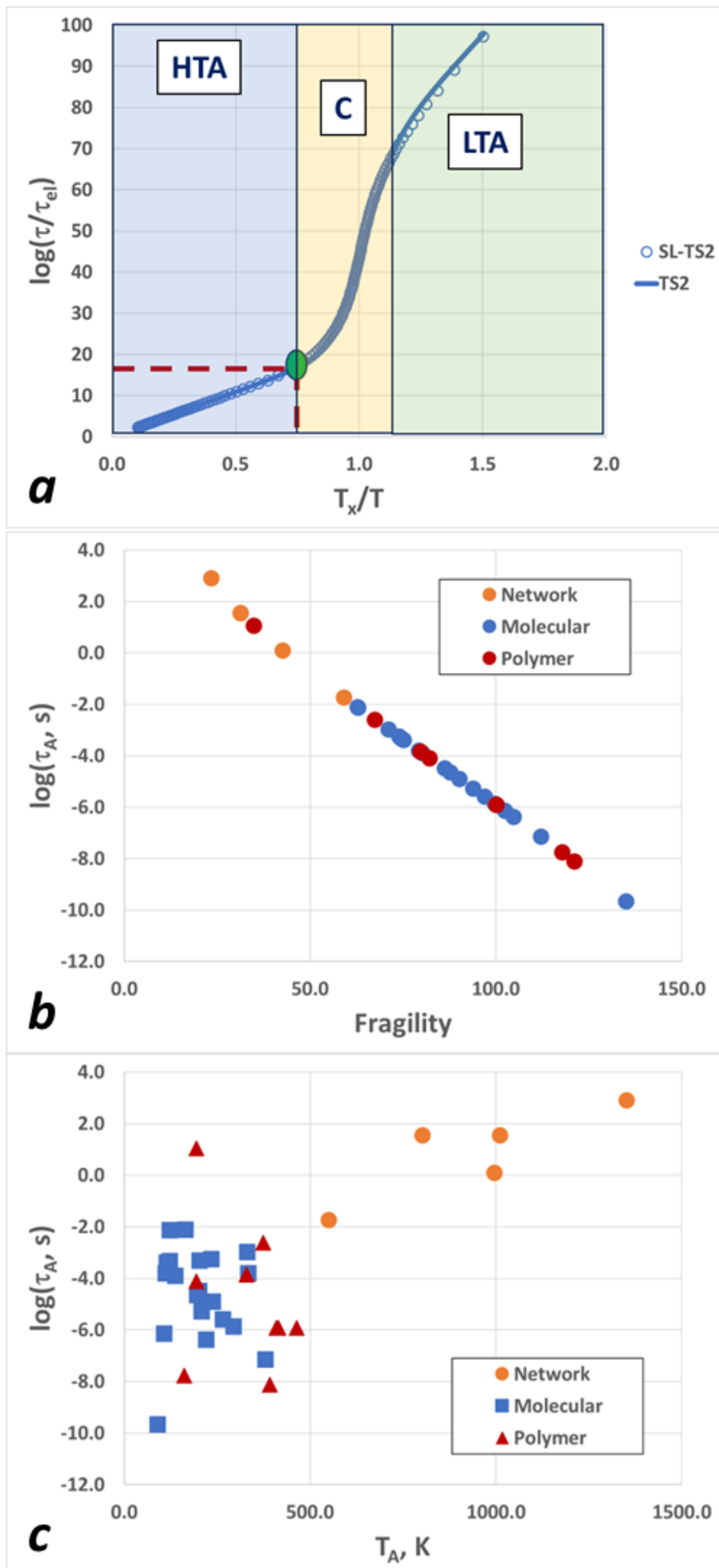

*Figure 3. (a) Arrhenius plot for the numerical SL-TS2 (circles) and analytical TS2 (solid blue line). The green circle and two red dashed lines describe the Arrhenius temperature and Arrhenius time, as described in the text. Abbreviations: HTA – high-temperature Arrhenius region, C – crossover; LTA – low-temperature Arrhenius region. (b) Logarithm of Arrhenius time vs. fragility for the 35 glass-formers; (b) Logarithm of Arrhenius time vs. Arrhenius temperature.*

*Arrhenius temperature and Arrhenius time.* The “mid-point temperature”, $T_x$, corresponds to the first-order phase transition between the “liquid” and “solid” states. However, for most materials, $T_x < T_g$; thus, the state of the material at $T = T_x$ is far from equilibrium in any real-life experiment. On the other hand, the glass transition temperature, $T_g$, does not have thermodynamic meaning – it is simply the point where the “Deborah number” (the ratio of the relaxation time and the experimental time [70]) approaches unity. One convenient way to introduce an accessible yet truly thermodynamic measure of temperature and time was proposed by Douglas and co-workers [71,72] and Simmons and co-workers. [73,74] They emphasized that upon cooling, the relaxation time exhibits a transition from Arrhenius to a super-Arrhenius (“crossover”) behavior, as shown in Figure 3a. Although the transition between the two regions is “fuzzy”, it is possible to define it algorithmically; for example, Simmons et al. [74] defined it as the point where $\log(\tau_\alpha)$ exceeds the Arrhenius extrapolation by 15%. In terms of TS2 or SL-TS2, this translates to the point where $\psi\left[\overline{T_A}\right] \equiv \psi_A \approx 0.05$. In other words, the ratio $\overline{T_A} = T_A/T_X \approx 1.23$ is material-independent, and so is the value of $\log(\tau_A/\tau_{el}) \approx 17.7$ (see the vertical and horizontal red lines in Figure 3a). It is thus straightforward to re-write equations (3) and (4) in terms of ratios $(T/T_A)$ and $\log(\tau_\alpha[T]/\tau_A)$, instead of $(T/T_X)$ and $\log(\tau_\alpha[T]/\tau_{el})$ (see Supplementary Information [66] for details).

We can investigate how the Arrhenius temperature and Arrhenius time relate to each other and to the dynamic fragility (Figures 3b and 3c). The logarithm of the Arrhenius time is a nearly perfect linear function of the fragility, with little to no scatter across families (Figure 3b). This is the consequence of the material-independent TS2 function we established earlier (see Supplementary Information [66] for its functional form). Such a behavior is reminiscent of the enthalpy-entropy or Meyer-Neldel compensation law. [75–78] It was recently shown [76] that the Arrhenius prefactor for the slow Arrhenius process (SAP) for various glass-forming liquids is not a universal parameter, but obeys the Meyer-Nelder mechanism;

we expect similar dynamics for the alpha process as well. The relationship between $\log(\tau_A)$ and $T_A$ (or $T_X$) is more complicated (Figure 3c). For high-temperature "network glasses" (silica, $B_2O_3$, Corning AS, and basalt), there is a clear trend reminiscent of an exponential behavior, $\tau_A(T_A) \cong \tau_A(0)\, exp[T_A/H_0]$, where $H_0$ is some enthalpy parameter. For the polymeric and molecular glasses, however, no such trend emerges and the data are completely scattered. At this time, we have to conclude that the two parameters setting the time and temperature scales are generally independent of each other.

*Debye-Waller factor and universality.* In their simulations, Simmons et al. [74] discovered that the Debye-Waller factor (mean-square displacement in the super-Arrhenius regime) can be normalized and rescaled to obey a simple temperature dependence,

$$\langle u^2 \rangle_r = \frac{3}{2} T_r - \frac{1}{2}, \qquad (6)$$

where $\langle u^2 \rangle_r \equiv \langle u^2(T) \rangle / \langle u^2(T_{ref}) \rangle$ and $T_r \equiv T/T_{ref}$ (see Figure 4a). In their study, Simmons et al. [74] chose $T_{ref} = T_A$; interestingly, we find that equation 6 holds for a relatively broad range of $T_{ref} > T_A$. Let us now consider the relationship between $log(\tau_r) \equiv log(\tau[T]/\tau[T_{ref}])$ and $\langle u^2 \rangle_r$. According to elastic activation models (Hall and Wolynes [79], Dyre [9,10], and others), one expects $\log(\tau_r) \propto (\langle u^2 \rangle_r)^{-1} - 1$. We attempt to find $T_{ref}$ for which this expression is valid and agrees with TS2 prediction. First, we re-cast the TS2 formula in terms of $T_r$ and $\tau_r$,

$$\log(\tau_r) = U_1 \left[\frac{1}{T_r} - 1\right] + (U_2 - U_1) \left[\frac{1}{T_r} \frac{1}{1 + \exp\left(V\left[\frac{1}{T_r} - \frac{T_{ref}}{T_x}\right]\right)} - \frac{1}{1 + \exp\left(V\left[1 - \frac{T_{ref}}{T_x}\right]\right)}\right] \qquad (7)$$

Here, $U_{1,2} = \left(\frac{1}{\ln(10)}\right) \overline{E_{1,2}} \left(\frac{T_x}{T_{ref}}\right)$, and $V = \left(\frac{\Delta S}{k_B}\right)\left(\frac{T_x}{T_{ref}}\right)$. Next, we use equation 6 to express $T_r$ as a function of $\langle u^2 \rangle_r$ and substitute into equation 7. Then, we can vary $T_{ref}$ and compute the dependence of $log(\tau_r)$ on $w = (\langle u^2 \rangle_r)^{-1} - 1$. In general, this dependence is highly nonlinear, but for $T_{ref} \approx 2.35 T_x$, it is well approximated by a straight line. (Note that $T_{ref} \approx 2.35 T_x$ is significantly higher than $T_A \approx 1.23 T_x$). Figure 4b shows the calculated TS2 line (red) together with a Hall-Wolynes straight line (blue) with the slope of 4.25 (± 0.1). Notably, the simulation data for various glass-formers collapse onto the TS2 or the Hall-Wolynes lines with no free parameters. In effect, given equation 6, this implies that the universal linear dependence of the Debye-Waller factor on the reduced temperature, combined with the material-independent TS2 relaxation time function, predicts the universal Hall-Wolynes elastic mechanism of local dynamics and relaxation.

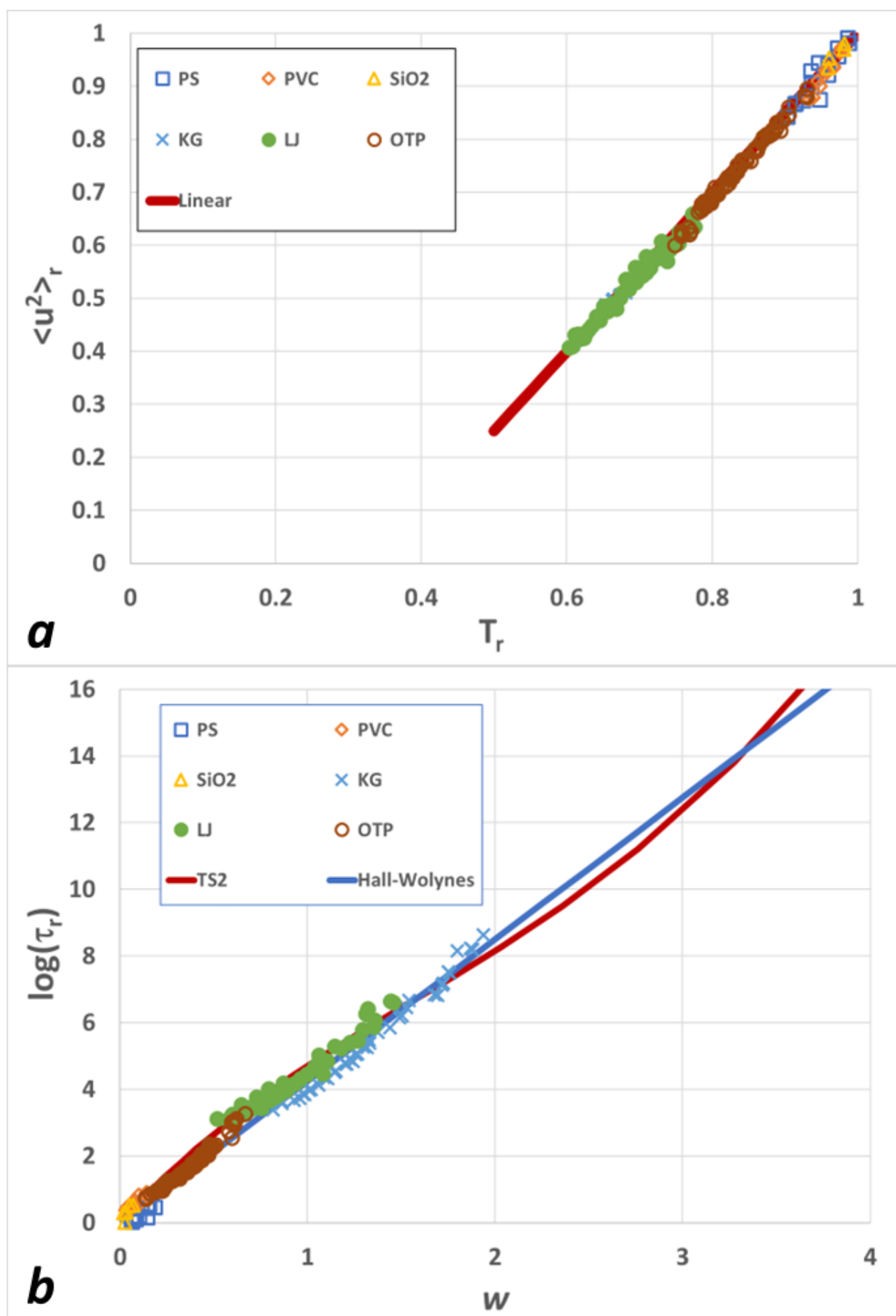

*Figure 4. (a) Normalized Debye-Waller factor vs. normalized temperature. Red line – TS2 function; symbols are representative systems from Ref. [74] as shown in the legend. (b) Logarithm of the normalized relaxation time vs. the shifted inverse Debye-Waller factor. Red line – TS2, blue line – elastic model, symbols are representative systems from Ref. [74]. Abbreviations: PS, PVC – all-atom OPLS simulations of 100-mer polystyrene and poly(vinyl chloride); LJ -- binary Lennard-Jones; KG -- Kremer-Grest polymer, SIO2 -- $SiO_2$ glass; OTP – ortho(terphenyl).*

*Conclusions.* We have demonstrated that the $\alpha$-relaxation behavior of multiple glass-formers (network, molecular, and polymer) can be successfully captured by a simple two-parameter model, where the two parameters are the "characteristic material temperature", $T_x$, and the "characteristic material time", $\tau_{el}$. The other, dimensionless, parameters of the master curve (the normalized high-temperature and low-temperature activation energies, $\frac{E_1}{k_B T_X} \approx 50(\pm 5)$ and $\frac{E_2}{k_B T_X} \approx 150(\pm 5)$, and the normalized entropy difference between the liquid and solid states, $\frac{\Delta S}{k_B} \approx 16(\pm 1)$) are broadly applicable to a variety of glassy systems, water being the only known exception so far. The knowledge of the relaxation time allows us to predict the nonequilibrium experiments like volume, enthalpy, and stress relaxation, or the specific volume and enthalpy changes upon heating or cooling. Longer-term, these results will help us develop new models for visco-elasto-plastic deformation in amorphous near-glassy materials. These will be subject of future studies.

*Acknowledgments.* RC acknowledges the support of the Office of Naval Research. VG thanks Faculty of Mechanical Engineering, Technion, for hospitality during his sabbatical. AZ gratefully acknowledges funding from the US Army DEVCOM Army Research Office through contract nr. W911NF-22-2-0256 and from the European Union through Horizon Europe ERC Grant number: 101043968 "Multimech". We thank Prof. David S. Simmons for providing simulation data, and Dr. Peter Lunkenheimer for helpful comments and suggestions.

The authors declare no competing financial interests.

Supporting Information for:

# General Two-Parameter Model of Alpha-Relaxation in Glasses

Valeriy V. Ginzburg[1], Oleg Gendelman[2], Riccardo Casalini[3], and Alessio Zaccone[4]

[1]Department of Chemical Engineering and Materials Science, Michigan State University, East Lansing, Michigan, USA 48824

[2]Faculty of Mechanical Engineering, Technion, Haifa 32000003, Israel

[3]Chemistry Division, Naval Research Laboratory, 4555 Overlook Avenue SW, Washington, D.C., USA 20375

[4]University of Milan, Department of Physics, via Celoria 16, 20133 Milano, Italy

## 1. Parameter table for the glass-formers

*Table S 1. Model parameters for the 35 glass-formers. See text for details*

| Material | $T_g$, K | $T_x/T_g$ | $T_x$, K | log($\tau_{el}$, s) | m | $T_A$, K | log($\tau_A$, s) | Ω |
|---|---|---|---|---|---|---|---|---|
| Silica | 1475 | 0.748 | 1104 | -14.8 | 23.5 | 1353 | 2.9 | 0.30 |
| Window Glass | 833 | 0.787 | 655 | -16.2 | 31.4 | 803 | 1.5 | 0.25 |
| Corning Aluminosilicate | 1050 | 0.787 | 826 | -16.2 | 31.4 | 1012 | 1.5 | 0.10 |
| Basalt | 993 | 0.819 | 813 | -17.6 | 42.7 | 996 | 0.1 | 0.24 |
| Glycerol | 191 | 0.869 | 166 | -21.0 | 74.5 | 203 | -3.3 | 0.18 |
| PDMS | 144 | 0.910 | 131 | -25.3 | 116.3 | 161 | -7.6 | 0.03 |
| PBD | 180 | 0.871 | 157 | -21.2 | 76.0 | 192 | -3.5 | 0.03 |
| PS | 373 | 0.890 | 332 | -23.0 | 93.9 | 407 | -5.3 | 0.13 |
| PMMA | 378 | 0.886 | 335 | -22.5 | 89.4 | 410 | -4.8 | 0.20 |
| PC | 423 | 0.883 | 373 | -22.3 | 86.7 | 458 | -4.6 | 0.13 |
| PVC | 350 | 0.904 | 316 | -24.5 | 108.8 | 388 | -6.8 | 0.36 |
| PVAc | 307 | 0.875 | 269 | -21.6 | 79.8 | 329 | -3.9 | 0.11 |
| PCHMA | 354 | 0.858 | 304 | -20.1 | 65.3 | 372 | -2.4 | 0.10 |
| PIB | 198 | 0.802 | 159 | -16.8 | 36.1 | 195 | 0.9 | 0.18 |
| 3MP | 80 | 0.925 | 74 | -27.4 | 135.2 | 91 | -9.7 | 0.64 |
| n-Propanol | 98 | 0.898 | 88 | -23.9 | 102.5 | 108 | -6.2 | 0.34 |
| EC | 104 | 0.875 | 91 | -21.5 | 79.4 | 112 | -3.8 | 0.38 |
| 3BP | 108 | 0.870 | 94 | -21.1 | 75.3 | 115 | -3.4 | 0.64 |
| EB | 115 | 0.870 | 100 | -21.1 | 74.6 | 123 | -3.3 | 0.57 |
| Toluene | 117 | 0.855 | 100 | -19.9 | 63.1 | 123 | -2.1 | 0.46 |
| Cumene | 129 | 0.876 | 113 | -21.6 | 80.3 | 138 | -3.9 | 0.56 |
| Propylene Carbonate | 158 | 0.854 | 135 | -19.8 | 62.9 | 165 | -2.1 | 0.37 |
| DP | 181 | 0.884 | 160 | -22.4 | 87.8 | 196 | -4.7 | 0.15 |
| MTD | 187 | 0.882 | 165 | -22.2 | 86.3 | 202 | -4.5 | 0.42 |
| DMP | 191 | 0.890 | 170 | -23.0 | 94.0 | 208 | -5.3 | 0.45 |
| TPP | 200 | 0.900 | 180 | -24.1 | 104.8 | 221 | -6.4 | 0.38 |
| OBP | 220 | 0.886 | 195 | -22.6 | 90.2 | 239 | -4.9 | 0.54 |
| Salol | 221 | 0.869 | 192 | -21.0 | 74.0 | 235 | -3.3 | 0.48 |
| OTP | 243 | 0.893 | 217 | -23.3 | 97.1 | 266 | -5.6 | 0.51 |
| Sorbitol | 268 | 0.896 | 240 | -23.6 | 99.8 | 294 | -5.9 | 0.18 |
| Glucose | 312 | 0.865 | 270 | -20.7 | 71.2 | 331 | -3.0 | 0.29 |
| CPDE | 312 | 0.875 | 273 | -21.5 | 79.4 | 335 | -3.8 | 0.27 |
| TNB | 342 | 0.906 | 310 | -24.9 | 112.2 | 380 | -7.2 | 0.37 |
| B2O3 | 530 | 0.849 | 450 | -19.5 | 59.2 | 551 | -1.7 | 0.32 |
| PDE | 298 | 0.883 | 263.2 | -22.3 | 87.1 | 323 | -4.6 | 0.21 |

Here, $T_g$ is the glass transition temperature, $T_x$ is the TS2 thermodynamic transition temperature, $\tau_{el}$ is the TS2 elementary time, *m* is the dynamic fragility, $T_A$ is the Arrhenius transition temperature as defined in the main text, $\tau_A$ is the relaxation time at $T = T_A$, and $\Omega$ is the “objective function” for the quality of the TS2 fit, defined as,

$$\Omega = \frac{1}{n}\sum_{k=1}^{n}\left|\log\left(\tau_{TS2}\left(T_k\right)\right) - \log\left(\tau_k\right)\right|$$

where $n$ is the number of datapoints, $T_k$ and $\tau_k$ are the temperature and the relaxation time for the datapoint $k$, and $\tau_{TS2}(T)$ is the TS2 relaxation time function. Given the fact that the $\alpha$-relaxation time is in reality the location of the maximum of the ε” function and considering other sources of error, the fit quality of 0.3 – 0.5 should be considered reasonably good.

## 2. Parameter table for the universal TS2 and SL-TS2 Models

*Table S 2. Parameters for the universal TS2 and SL-TS2 models.*

| **TS2 Parameters** | |
|---|---|
| $E_1/RT_x$ | **50 (±5)** |
| $E_2/RT_x$ | **150 (±10)** |
| $\Delta S/R$ | **16 (±1)** |
| | |
| **SL-TS2 Parameters** | |
| $\alpha_{LL}$ | **0.934 (±0.005)** |
| $r_S$ | **388 (±3)** |
| $r_L$ | **412 (±3)** |

### 3. Shifted log-log plots for the $\alpha$-relaxation time vs. temperature for various molecular glasses

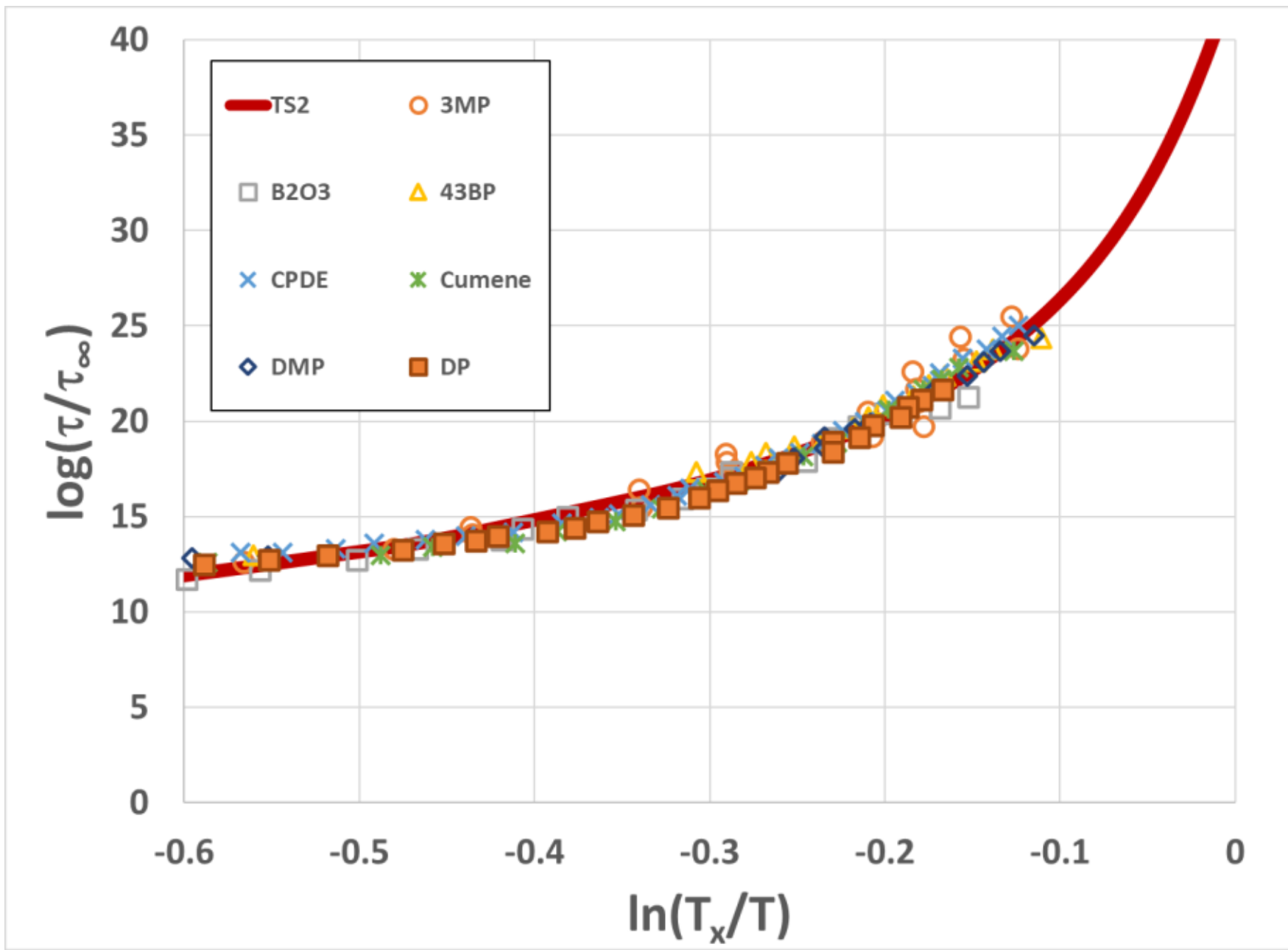

*Figure S 1. Log-log plot of scaled relaxation time vs. scaled temperature for a subset of molecular glass-formers. TS2 is the universal TS2 model curve, B2O3 is boron oxide, 3MP is 3-methyl pentane; 43BP is 4,3-bromopentane; CPDE is cresolphthalein-dimethylether; DMP is dimethyl phthalate; DP is diethyl phthalate.*

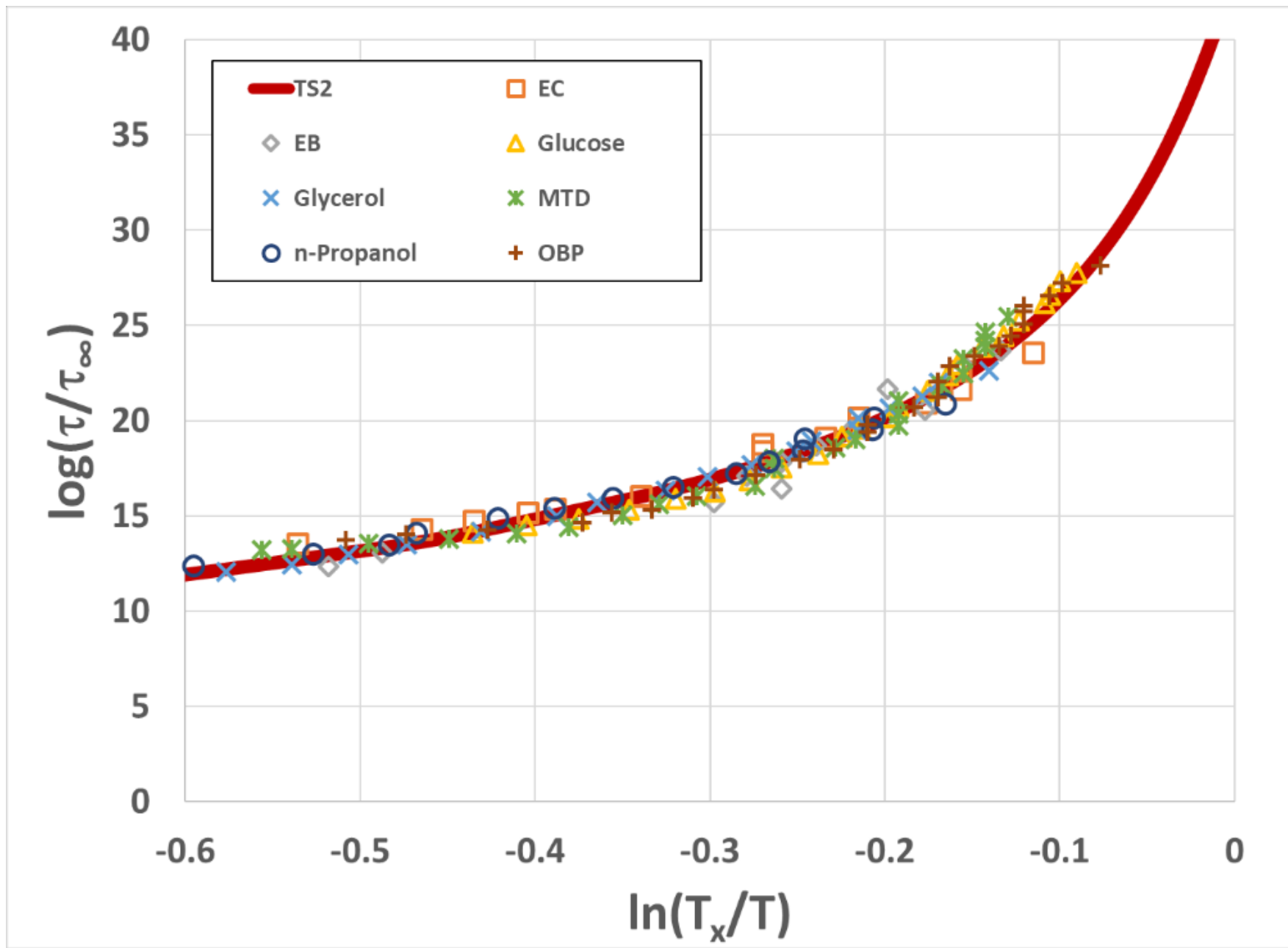

*Figure S 2. Same as Figure S1, but with a different subset of molecular glasses. Abbreviations: EC – ethyl cyclohexane; EB is ethylbenzene; MTD is m-toluidine; OBP is o-benzylphenol, also known as α-phenyl-o-cresol.*

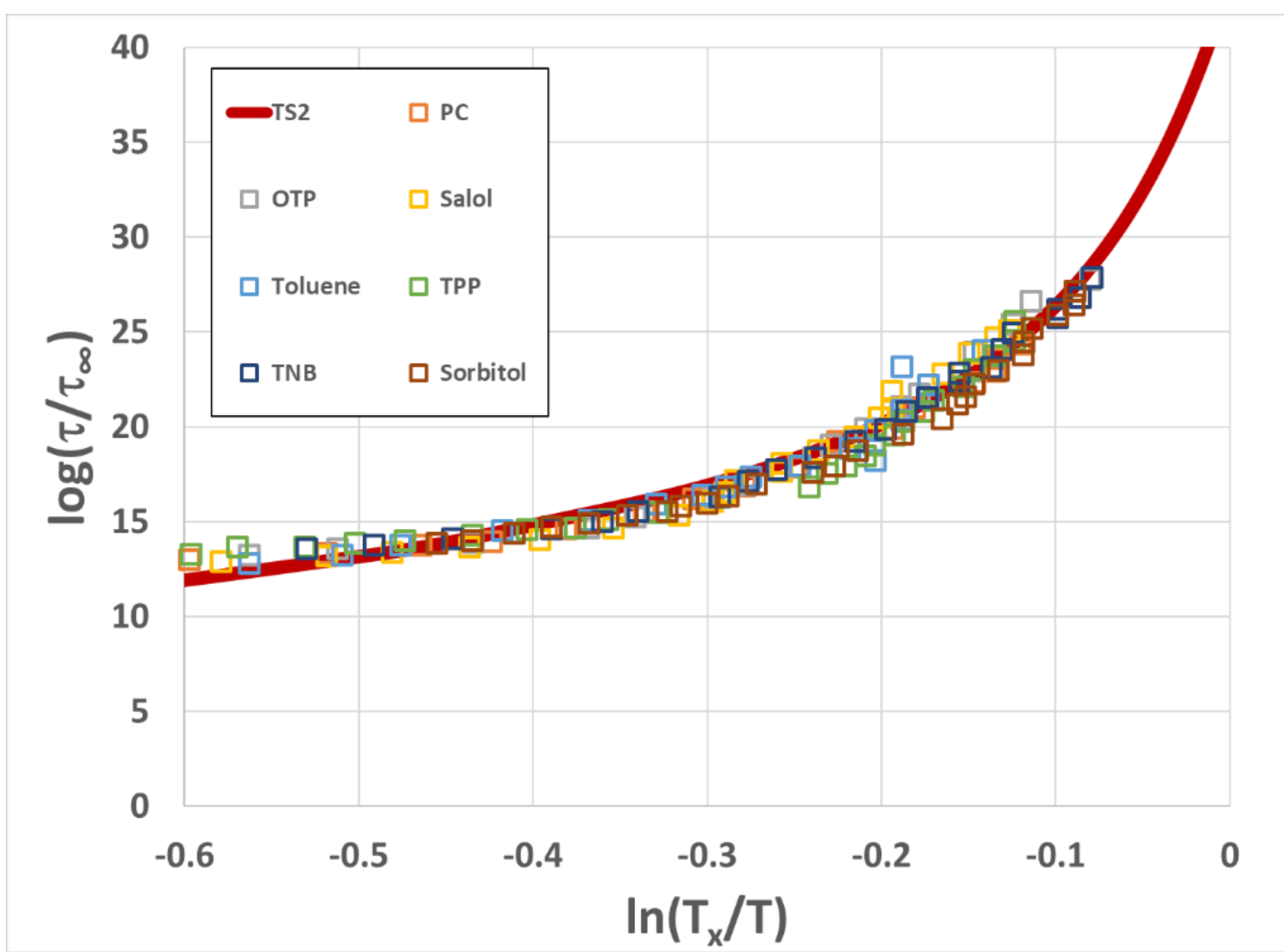

*Figure S 3. Same as Figures S1 – S2, but with a different subset of molecular glasses. Abbreviations: PC – propylene carbonate; OTP – ortho-terphenyl; TPP -- triphenyl phosphite; TNB -- tri- $\alpha$-naphtyl benzene.*

## 4. Relationship between fragility and Arrhenius relaxation time

The fragility within TS2 can be evaluated as,

$$m = \frac{\partial \log(\tau_\alpha)}{\partial\left[\frac{T_g}{T}\right]_{T=T_g}} = \frac{1}{\ln(10)}\frac{T_x}{T_g}\left[\overline{E_1} + \left(\overline{E_2} - \overline{E_1}\right)\psi_g + \frac{T_x}{T_g}\left(\overline{E_2} - \overline{E_1}\right)S\psi_g\left(1-\psi_g\right)\right] \quad \text{(S1)}$$

Where $S \equiv \Delta S / R$, and

$$\psi_g = \left[1 + \exp\left(S\left[\frac{T_x}{T_g} - 1\right]\right)\right]^{-1} \quad \text{(S2)}$$

The relaxation time at T = $T_A$ is given by,

$$\log\left(\frac{\tau_g}{\tau_A}\right) = \frac{1}{\ln(10)}\left[\overline{E_1}\left(\frac{T_X}{T_g} - \frac{T_X}{T_A}\right) + \left(\overline{E_2} - \overline{E_1}\right)\left(\frac{T_X}{T_g}\psi_g - \frac{T_X}{T_A}\psi_A\right)\right] \quad \text{(S3)}$$

As discussed in the main manuscript, the ratio $T_x/T_A$ is a universal constant once we have chosen $\psi_A = 0.05$; $T_x/T_A = Q = 0.816$. Recalling that $\log(\tau_g) = 2$, we can express $\log(\tau_A)$ as a function of a single unknown, $T_x/T_g$,

$$\log(\tau_A) = 2 - \frac{1}{\ln(10)}\left[\overline{E_1}\left(\frac{T_X}{T_g} - Q\right) + \left(\overline{E_2} - \overline{E_1}\right)\left(\frac{T_X}{T_g}\psi_g - Q\psi_A\right)\right] \quad \text{(S4)}$$

Note that $\psi_g$ is given by equation S2 and is thus also a function of $T_x/T_g$. We can solve equation S4 numerically to express $T_x/T_g$ as a function of $\log(\tau_A)$, and then substitute into equation S1 to determine the relationship between $m$ and $\log(\tau_A)$, as plotted in Figure 3b of the main manuscript. Below, in Figure S4, we re-plot the data from Figure 3b and fit $\log(\tau_A)$ vs. $m$ using a fifth-order polynomial function.

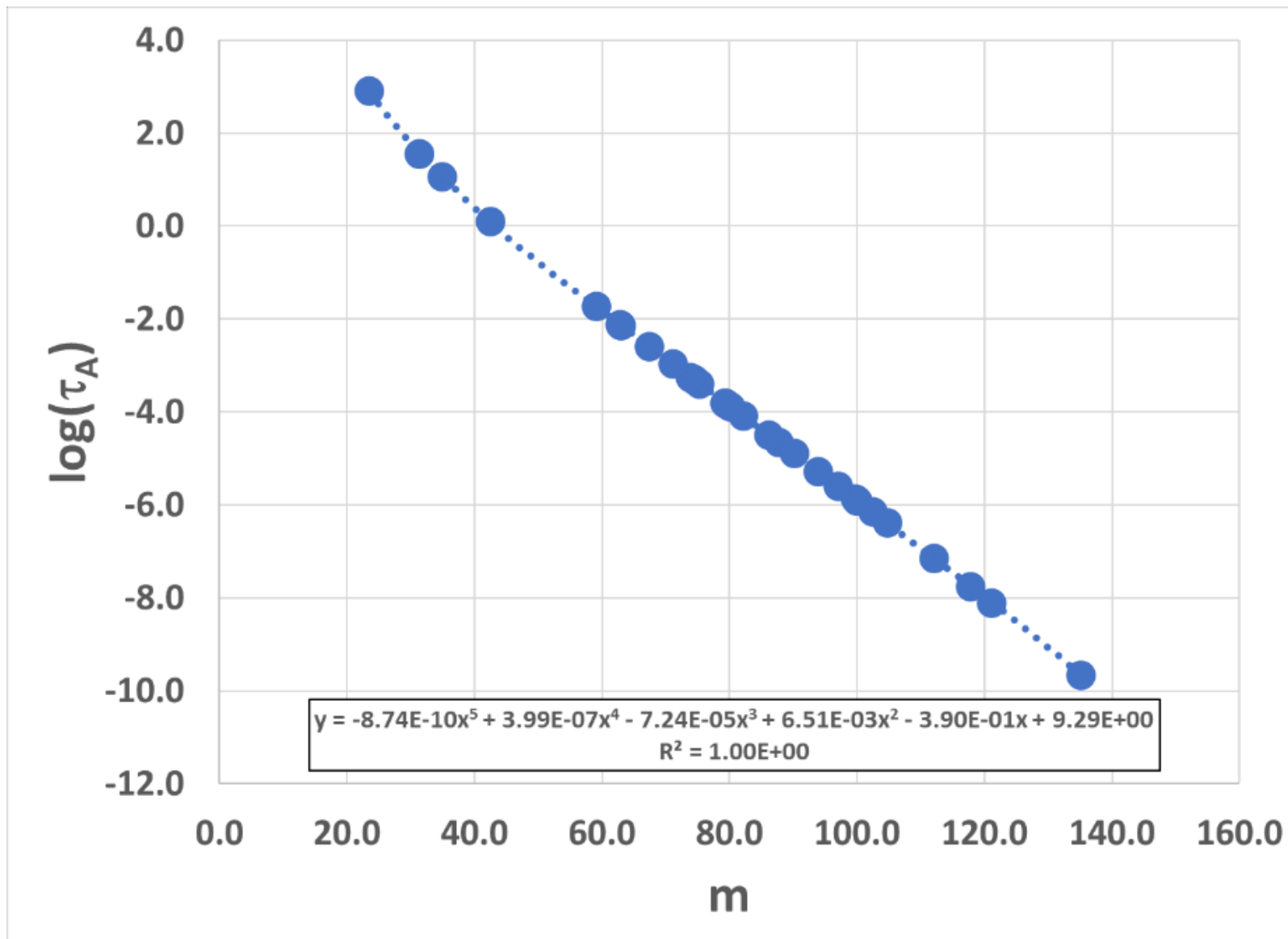

*Figure S 4. Logarithm of the Arrhenius relaxation time as a function of fragility – the data for all 34 glass-formers and the fifth-order polynomial fit.*

## 5. The Wang-Porter estimate of the liquid-state activation energy

Wang and Porter [1] derived the following expression for the activation energy in the liquid state of amorphous polymers (equation 11 of their manuscript),

$$E^* = (R/\alpha_L)(T_g + M)^2 / \left((0.23T_g + M)^2\right) \quad \text{(S5)}$$

Here, $R$ is the gas constant, $\alpha_L$ is the coefficient of thermal expansion above $T_g$, and $M = T_A - T_g$. We can rewrite this equation to obtain an expression for $\overline{E}_1$,

$$\overline{E}_1 \equiv \frac{E^*}{RT_X} = \left(\frac{T_g}{T_X}\right)\left(\frac{1}{\alpha_L T_g}\right)\left(\frac{1+\left[\frac{T_A}{T_g}-1\right]}{0.23+\left[\frac{T_A}{T_g}-1\right]}\right)^2 \quad \text{(S6)}$$

Estimating $\left[\frac{T_A}{T_g}-1\right] \approx 0.2$ [2], $\alpha_L T_g \approx 0.195 \pm 0.03$ Simha [3], and $\frac{T_X}{T_g} \approx 0.8$ (see the main manuscript), we obtain $\overline{E}_1$=*52 ± 7*, in a perfect agreement with the regressed value $\overline{E}_1$=50 ± 5.

## 6. Functional form for TS2 relaxation time relative to the Arrhenius time and Arrhenius temperature

The universal TS2 relaxation time as a function of temperature is given by,

$$\log\left(\frac{\tau_\alpha[X]}{\tau_\infty}\right) = \frac{1}{\ln(10)}\left[\frac{\overline{E}_1}{X} + \frac{\overline{E}_2 - \overline{E}_1}{X}\psi[X]\right] \tag{S7}$$

Where $X = T/T_X$ and $\psi[X] = \left\{1 + \exp\left[\frac{\Delta S}{R}\left(1 - X^{-1}\right)\right]\right\}^{-1}$. Given the definition of the Arrhenius temperature as the temperature for which ψ = $\psi_A$ = 0.05, we can easily find $X_A = T_A/T_X$,

$$X_A = 1 - \frac{1}{S}\ln\left(\frac{1}{\psi_A} - 1\right) \approx 0.816 \tag{S8}$$

Thus,

$$\log\left(\frac{\tau_\alpha[Z]}{\tau_A}\right) = \frac{1}{\ln(10)}\left[\frac{\overline{E}_1}{X_A}\left(Z^{-1} - 1\right) + \frac{\overline{E}_2 - \overline{E}_1}{X_A}\left(\psi[Z]Z^{-1} - \psi_A\right)\right] \tag{S9}$$

Where $Z = T/T_A$.

## 7. Residuals

Below, we plot residuals, $y_k = \log\left(\tau_{TS2}\left(T_k\right)\right) - \log\left(\tau_k\right)$, for the glass-formers studied here. The figures indicate that the residuals are distributed reasonably randomly around the horizontal axis, indicating, once again, a reasonable quality of the TS2 fit.

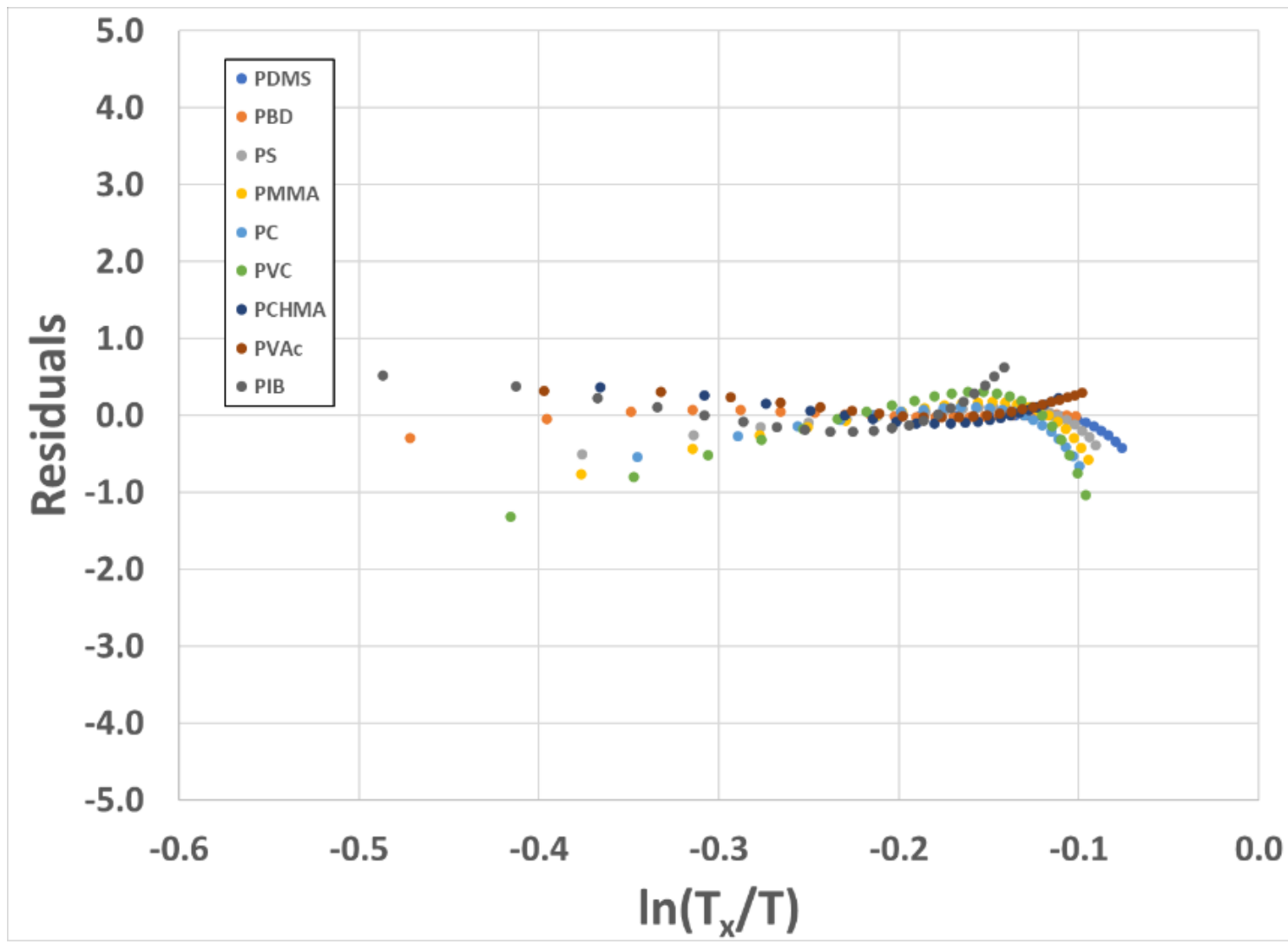

*Figure S 5. Residuals for the polymer glass-formers*

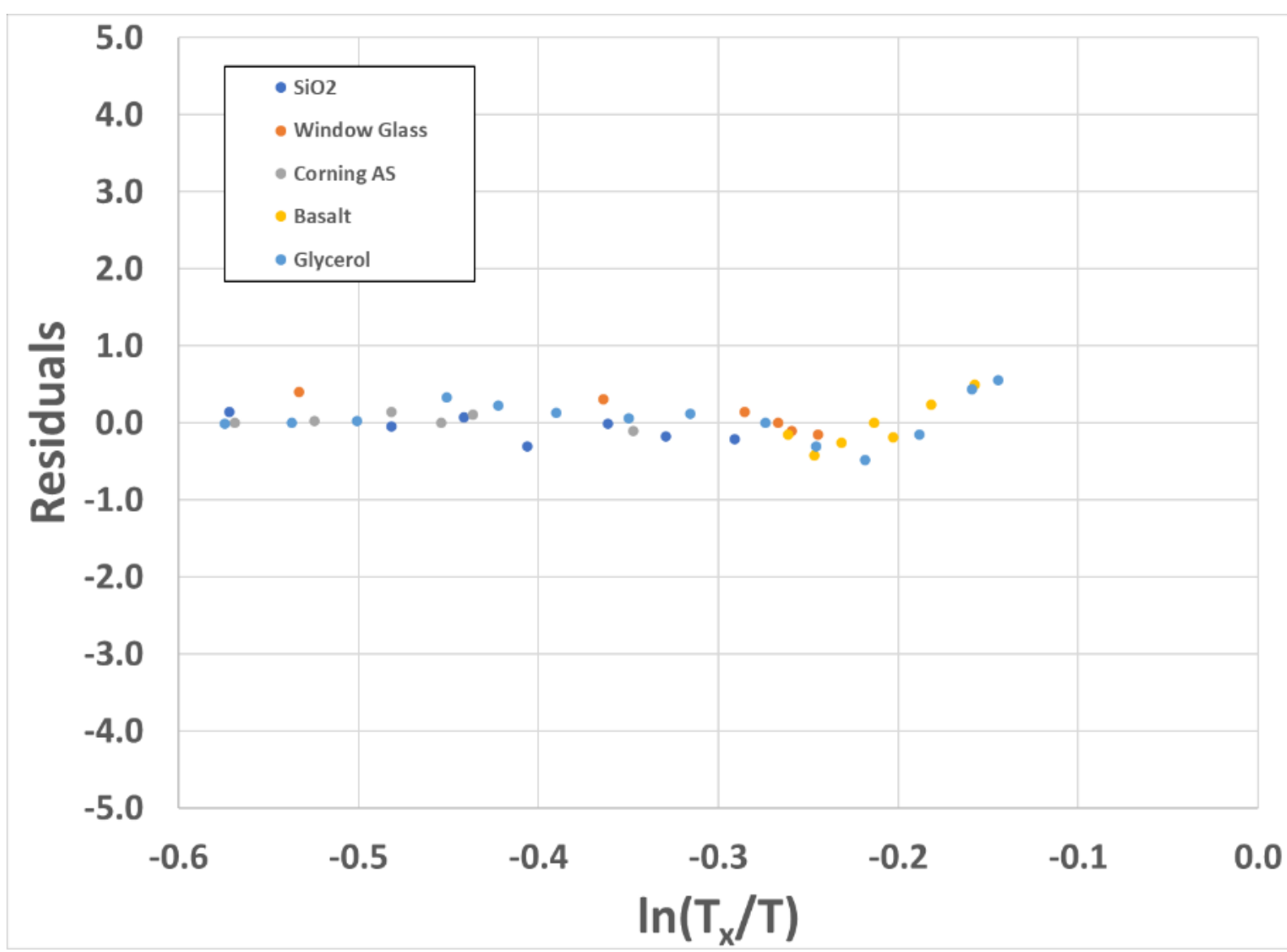

*Figure S 6. Residuals for network and some molecular glass-formers*

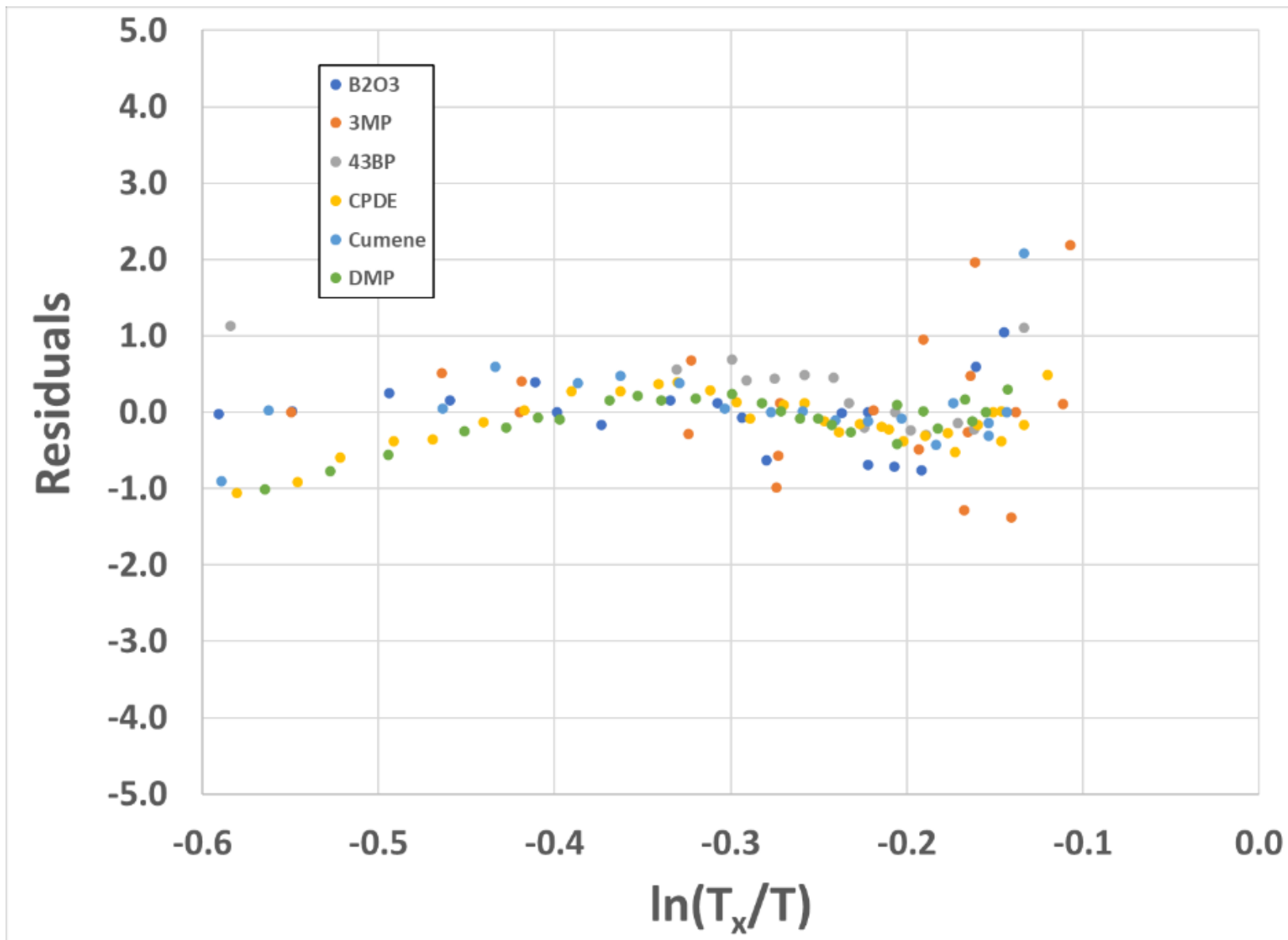

*Figure S 7. Residuals for the glass-former set depicted in Figure S1*

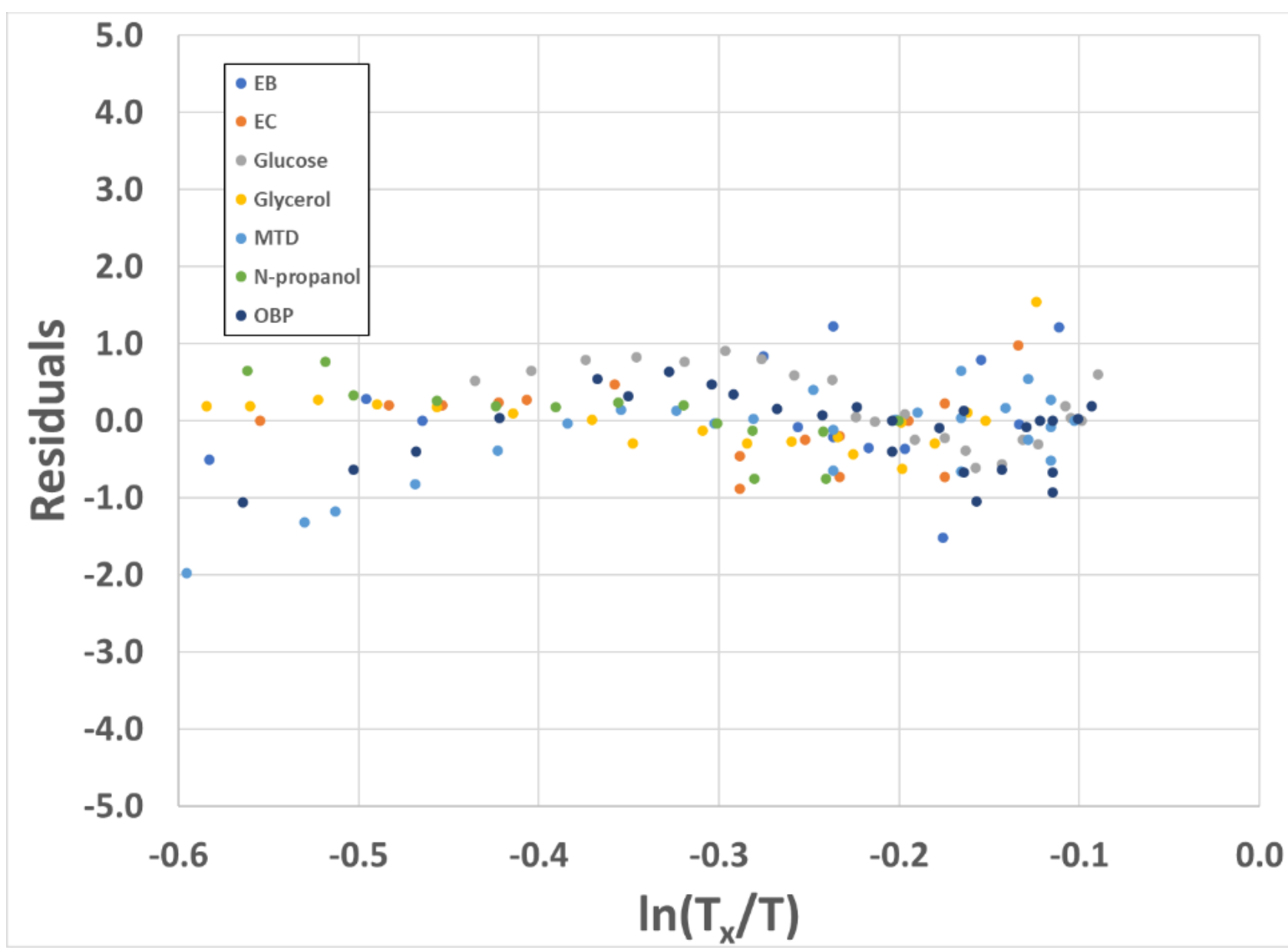

*Figure S 8. Residuals for the glass-former set depicted in Figure S2*

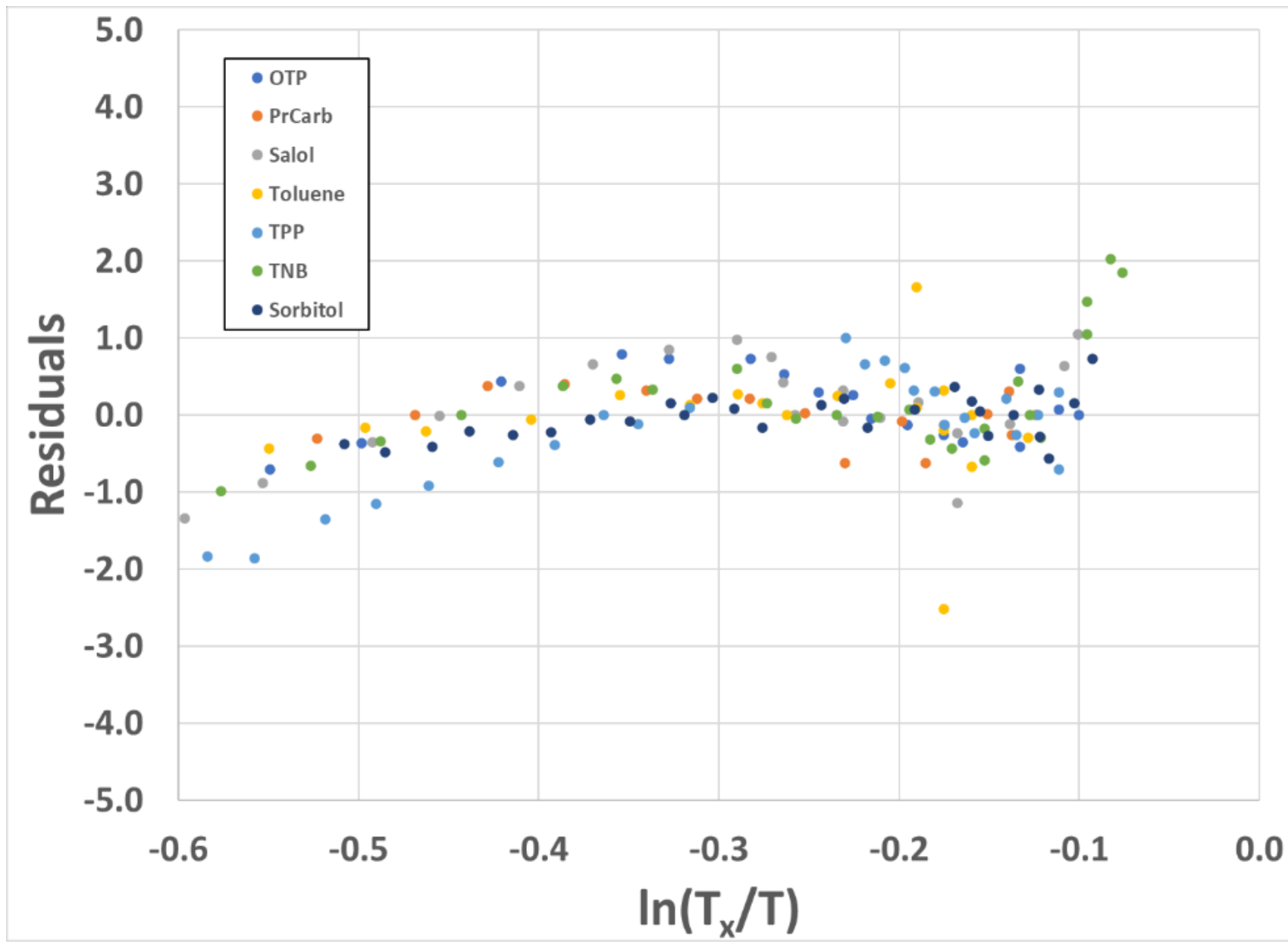

*Figure S 9. Residuals for the glass-former set depicted in Figure S3*